\documentstyle[12pt]{article}

\newcommand{\be}{\begin{equation}}
\newcommand{\ee}{\end{equation}}
\newcommand{\sS}{\stackrel{\rightarrow}{S}}
\newcommand{\sB}{\stackrel{\rightarrow}{B}}
\newcommand{\se}{\stackrel{\rightarrow}{e}}
\newcommand{\sr}{\stackrel{\rightarrow}{r}}
\newcommand{\sk}{\stackrel{\rightarrow}{k}}
\newcommand{\sq}{\stackrel{\rightarrow}{q}}
\newcommand{\bt}{\beta}
\newcommand{\om}{\omega}
\newcommand{\lmb}{\lambda}
\newcommand{\gm}{\gamma}
\newcommand{\ep}{\varepsilon}
\newcommand{\sgm}{\Sigma}
\newcommand{\prt}{\partial}
\newcommand{\lgl}{\langle}
\newcommand{\rgl}{\rangle}
\newcommand{\ra}{\rightarrow}

\begin{document}

\begin{center}

{\Large{\bf Interplay between Mesoscopic and Microscopic Fluctuations in
Ferromagnets} \\ [5mm]

V.I. Yukalov} \\ [3mm]

{\it Centre for Interdisciplinary Studies in Chemical Physics \\
University of Western Ontario, London, Ontario N6A 3K7, Canada \\
and \\
Bogolubov Laboratory of Theoretical Physics \\
Joint Institute for Nuclear Research, Dubna 141980, Russia}

\end{center}

\vspace{8cm}


\vspace{2cm}

{\bf keywords:} mesoscopic fluctuations, heterophase ferromagnets, phase
transitions

\newpage

\begin{abstract}

A model of a ferromagnet is considered, in which there arise mesoscopic
fluctuations of paramagnetic phase. The presence of these fluctuations
diminishes the magnetization of the ferromagnet, softens the spin--wave
spectrum, increases the spin--wave attenuation, shortens the magnon free
path, lowers the critical point, and can change the order of phase
transition. A special attention is paid to the interplay between these
mesoscopic paramagnetic fluctuations and microscopic fluctuations due to
magnons. One of the main results of this interplay is an essential
extension of the region of parameters where the ferromagnet--paramagnet
phase transition is of first order.

\end{abstract}

\newpage

\section{Introduction}

When one introduces into a model of a ferromagnet additional variables,
sometimes called hidden, such as phonons or impurities, the behaviour of 
the model can be essentially distorted. Two main types of changes arise: A
renormalization of critical exponents of a continuous transition or the
disruption of a continuous transition to the first order one [1,2] (and
literature therein). Such changes are caused by the interplay of different
fluctuations in a ferromagnet, say, due to magnons and to phonons.

In the present paper we study the interplay between two types of
fluctuations in ferromagnets, which has not been examined earlier. Let us
consider a ferromagnet in which fluctuations of paramagnetic phase can
occur. This kind of fluctuations is called {\it heterophase} [3]. Such
fluctuations are necessarily of {\it mesoscopic} nature, either in space
or in time or both. Mesoscopic in space means that their characteristic
size is much larger than the average interparticle distance but much
smaller than the size of a system. Mesoscopic in time implies that their
typical lifetime is much longer than the characteristic time of an
elementary oscillation but much shorter than the observation time. An
elementary oscillation, in the case of spin systems, concerns fast spin
motions whose typical time can be evaluated as $\tau_{loc}\approx\hbar/J$,
where $\hbar$ is the Planck constant and $J$ is an exchange integral.
Taking $J\sim(10^{-12}-10^{-13})\; erg$, we have $\tau_{loc}\sim(10^{-14}-
10^{-13})\; s$. Since heterophase fluctuations are always mesoscopic, we
may combine these two notions into one adjective mesophase. Thus, {\it
mesophase fluctuations} are, by definition, mesoscopic fluctuations of
one phase inside a host phase.

Ferromagnets with such mesoscopic fluctuations have been considered
earlier [3--5] in the mean--field approximation. The latter takes into
account only flippon states [3] of spin systems. However, in real magnets
there are also magnon excitations. These excitations describe small
oscillations of spins around their equilibrium positions and, thus, are to
be treated as {\it microscopic fluctuations}.

The aim of this paper is to analyse the properties of a ferromagnet in
which there occur both types of fluctuations discussed above, microscopic
magnon fluctuations and mesoscopic paramagnetic fluctuations. There are
many examples of heterophase magnets (see review [3]) whose mesoscopic
heterogeneous structure can be confirmed by neutron scattering experiments, 
such as diffuse scattering, small--angle scattering, Bragg reflections,
and polarized--neutron scattering [6]. This is why it is important to
study the properties of mesophase magnets in a more realistic, than
mean--field, approximation.  Note also that a system with mesophase
fluctuations is nothing but another name for a system with mesoscopic
phase separation that is now intensively studied in connection with
high--temperature superconductors [7--11].

\section{Mesophase Ferromagnet}

Consider a ferromagnet in which there appear mesoscopic paramagnetic
fluctuations [12]. Such a system is nonuniform and quasiequilibrium. The
general theory of statistical systems with mesoscopic fluctuations is
developed in Ref.[3]. Following this theory, we average over all possible
configurations of heterophase fluctuations coming to an effective
Hamiltonian
\be
H_{eff} =H_1\oplus H_2
\ee
where $H_\nu$, with $\nu=1,2$, is a phase--replica Hamiltonian
representing a pure ferromagnetic ($\nu=1$) or paramagnetic ($\nu=2$)
phase. Each $H_\nu$ is defined on a weighted space ${\cal H}_\nu$ of
quantum states typical of the corresponding phase [3]. The effective
Hamiltonian (1) acts on a yield--weighted fiber space
\be
{\cal Y} = {\cal H}_1\otimes {\cal H}_2 \; .
\ee
A phase--replica Hamiltonian $H_\nu$, for the Heisenberg--type spin
interactions, reads
\be
H_\nu = \frac{1}{2} w_\nu^2 NU - w_\nu^2\sum_{i\ne j}J_{ij}
\sS_{i\nu} \cdot \sS_{j\nu} - w_\nu\sum_i \sB \cdot \sS_{i\nu} \; ,
\ee
where $N$ is the number of lattice sites; $U$, a crystal--field parameter;
$J_{ij}=J_{ji} > 0$, an exchange interaction; $\sS_{i\nu}$, a spin
operator defined on the weighted space ${\cal H}_\nu$, so that
$\sS_{i\nu}{\cal H}_\nu = \sS_i{\cal H}_\nu;\; i,j=1,2,\ldots, N$;
\be
\sB =\mu_0 H_0\se_z \equiv B\se_z \; ,
\ee
$\mu_0$ being a particle magnetic moment; $H_0$, external magnetic field
in the $z$--direction. The factors $w_\nu$ are the geometric phase
probabilities satisfying the conditions
\be
0 \leq w_\nu \leq 1 \; , \qquad w_1 + w_2 = 1\; ,
\ee
and defined by minimizing the effective free energy
\be
f = -\frac{T}{N}\ln\; Tr\exp\left ( -\bt H_{eff}\right ) \; ,
\ee
in which $T$ is temperature, $\bt T\equiv 1$, and the trace is taken over
the fiber space (2). With the notation
$$
w\equiv w_1 \; , \qquad w_2 \equiv 1 - w \; ,
$$
we have
\be
\frac{\prt f}{\prt w} = 
\frac{1}{N}\lgl \frac{\prt}{\prt w} H_{eff}\rgl = 0 \; ,
\ee
where $\lgl\ldots\rgl$ implies a statistical average with the statistical
operator
$$
\hat\rho = \frac{\exp(-\bt H_{eff})}{Tr\;\exp(-\bt H_{eff})} \; .
$$

The order parameters
\be
S_\nu \equiv \frac{1}{N}\sum_{i=1}^N\lgl S_{i\nu}^z\rgl \qquad
(\nu = 1,2)
\ee
are defined so that for the ferromagnetic phase
\be
\lim_{B\ra 0}\lim_{N\ra\infty} S_1 \not\equiv 0 \; ,
\ee
while for the paramagnetic phase
\be
\lim_{B\ra 0}\lim_{N\ra\infty} S_2 \equiv 0 \; .
\ee

Introducing the notation
\be
\Sigma_\nu \equiv \frac{1}{N}\sum_{i\neq j} J_{ij}\lgl \sS_{i\nu}
\cdot \sS_{j\nu}\rgl \; ,
\ee
we obtain from (7) an equation
\be
w = \frac{U - 2\Sigma_2 + B( S_1 - S_2)}{2( U - \Sigma_1 - \Sigma_2)}
\ee
for the ferromagnetic--phase probability. Here it is assumed that 
$U\neq\sgm_1+\sgm_2$. In the case when the latter assumption does not
hold, the equation for $w$ reads as
$$
\sgm_1 - \sgm_2 + B ( S_1 - S_2 ) = 0 \qquad (U = \sgm_1 + \sgm_2) \; .
$$

In order that (12) would satisfy the probability property $0\leq w\leq 1$,
one of the following sets of inequalities is to be valid, either
$$
U > \sgm_1 + \sgm_2 \; ,
$$
\be
U \geq 2\sgm_1 + B( S_1 - S_2 ) \; ,
\ee
$$
U \geq 2\sgm_2 - B( S_1 - S_2 ) \; ,
$$
or
$$
U < \sgm_1 + \sgm_2 \; ,
$$
\be
U \leq 2\sgm_1 + B( S_1 - S_2 ) \; ,
\ee
$$
U \leq 2\sgm_2 - B( S_1 - S_2 ) \; .
$$
The magnetization of the ferromagnetic phase is expected to be larger
than that of the paramagnetic phase, i.e., $S_1>S_2$. Keeping this in mind, 
we may notice from (12) that external magnetic field supresses paramagnetic 
fluctuations, that is, $w_2\ra 0$ as $B$ increases. Paramagnetic fluctuations
disappear at the critical field
$$ 
B_c = \frac{U - 2\sgm_1}{S_1 - S_2} \qquad (w=1) \; ,
$$
when $w_2=0$.

Equation (12) for the ferromagnetic probability $w$ resulted from the
extremum condition (7). In order that this extremum be a minimum, the
inequality 
$$
\frac{\prt^2f}{\prt w^2} >0
$$ 
must hold. Calculating the second derivative of (6), we take into account
(7) and the equality
$$
\frac{\prt\rho}{\prt w} = -\bt\rho\; \frac{\prt H_{eff}}{\prt w} \; .
$$
Then we find
$$
\frac{\prt^2f}{\prt w^2} =\frac{1}{N} \lgl
\frac{\prt^2H_{eff}}{\prt w^2} \rgl - \frac{\bt}{N}\lgl\left (
\frac{\prt H_{eff}}{\prt w}\right )^2 \rgl \; .
$$
The second term in the right--hand side of the latter equality is 
positively defined (nonnegative). Hence a {\it necessary} condition for
$\prt^2f/\prt w^2 >0$ is the positiveness of the first term, which gives
\be
U > \sgm_1 + \sgm_2 \; .
\ee
Therefore, between two sets (13) and (14) we must choose the set of
inequalities (13). The set (14) can correspond only to a metastable state.

\section{Spin--Wave Approximation}

Consider, first, the low temperatures
\be
T \ll J \equiv \frac{1}{N} \sum_{i\neq j} J_{ij} \; ,
\ee
when the Dyson [13,14] spin--wave approximation provides accurate
asymptotic expansions. Then for the free energy (6) of a mesophase
ferromagnet, with spin one half and for zero external field, one has
\be
f \simeq \left ( w^2 - w + \frac{1}{2}\right ) U - \frac{w^2}{4} J -
\frac{T}{w^3\rho a_0^3} \left ( \frac{T}{2\pi J}\right )^{3/2}
\left [ \zeta\left (\frac{5}{2}\right ) +
\frac{9T}{8w^2 J}\zeta\left (\frac{7}{2}\right )\right ] \; ,
\ee
where $\rho$ is a spin density, $\zeta(\cdot)$ is the Riemann zeta
function, and
$$
a_0^2 \equiv \frac{1}{N}\sum_{i\neq j} \sr^2_{ij}\frac{J_{ij}}{3J} \; ,
\qquad \sr_{ij} \equiv \sr_i - \sr_j \; .
$$

For the average magnetization
$$
M \equiv \mu_0 \left ( w_1 S_1 + w_2 S_2 \right ) \; ,
$$
in the zero field $B=0$, when $S_2=0$, we find
\be
\frac{M}{\mu_0} \simeq \frac{w}{2} - 
\frac{\zeta\left (\frac{3}{2}\right )}{w^2\rho a_0^3}
\left ( \frac{T}{2\pi J}\right )^{3/2} \; ,
\ee
which shows that, due to disordering caused by mesoscopic fluctuations,
the low--temperature magnetization diminishes.

The spin--wave spectrum [15--17], in the presence of mesophase
fluctuations softens, as is seen from the expression
$$
\ep(\sk) = 2 w^2 C \left [ J - J(\sk) \right ] -
$$
\be
- \frac{2w^2C}{(2\pi)^3\rho} \int
\frac{J - J(\sk) + J(\sk +\sq) - J(\sq)}
{\exp\left ( w^2JCa_0^2q^2/ T\right ) - 1} d\sq \; ,
\ee
in which
$$
C \equiv S_1 \; , \qquad J(0) = J \; , 
$$
$$ 
J(\sk) =\frac{1}{N}\sum_{i\neq j} J_{ij} e^{-i\sk\sr_{ij}} \; , 
$$
$$
J_{ij} =\frac{1}{(2\pi)^3\rho}\int J(\sk) e^{i\sk\sr_{ij}}d\sk \; ,
$$
and the integration goes over the Brillouin zone [18]. In the long--wave
limit, the spin--wave spectrum (19) yields
$$
\ep(\sk) \simeq w^2 J C a_0^2 k^2 \qquad (k\ra 0) \; .
$$

The existence of mesophase fluctuations enhances the attenuation
$\gamma(k)$ of spin waves due to magnon--magnon scattering and shortens
the magnon free path $\lmb(k)$, according to the formulas
$$
\gm(k) \simeq \frac{J a_0^2k^3}{2w^3}\left (\frac{2T}{J}\right )^{5/2} 
\qquad \left ( T \ll \ep(\sk) \right ) \; ,
$$
\be
\lmb(k) \simeq \frac{w^5 a_0^5\rho^2}{\zeta(3/2)k^2} \left (
\frac{\pi J}{T}\right )^{5/2} \qquad ( T \ll J ) \; ,
\ee
which are obtained using the techniques of Refs. [13,14,19]. The
shortening of the magnon free path is naturally related to the decrease of
the magnon lifetime resulting from the increase of attenuation.

\section{Mean--Field Approximation}

The model (1)--(3) of a mesophase ferromagnet in the mean--field
approximation has been analysed earlier [3--5]. Here we recall only some
main formulas and consider the influence of mesophase fluctuations on the
phase transition order. The consideration below deals again with spin one
half and zero external field. For convenience, we introduce the notation
\be
t \equiv\frac{T}{J} \; , \qquad
u\equiv \frac{U}{J} \; .
\ee

For the average spin $C\equiv S_1$, we have
\be
C =\frac{1}{2}{\rm tanh}\frac{w^2C}{t} \; .
\ee
The ferromagnetic--phase probability (12) becomes
\be
w = \frac{u}{2(u-C^2)} \; .
\ee
From (22) and (23) we may get
\be
t = u^2 C \left [ 2(u - C^2)^2 \ln\frac{1+2C}{1-2C}\right ]^{-1} \; .
\ee
In the vicinity of the critical point, when $C\ll 1$, expanding the
right--hand side of (24) in powers of $C$, we find
$$
8t \simeq 1 +\left ( \frac{2}{u} - \frac{4}{3}\right ) C^2 +
\left ( \frac{3}{u^2} - \frac{8}{u} + 2.133 \right ) C^4 +
$$
\be
+ \left ( \frac{4}{u^3} - \frac{28}{u^2} +\frac{21.333}{u} 
- 3.454\right ) C^6 \; .
\ee
This gives either
\be
C \simeq \sqrt{\frac{4u}{2u-3}}\; \left ( t_c - t\right )^{1/2}  \qquad
\left ( u\neq \frac{3}{2}\right ) \; ,
\ee
where
$$
t \ra t_c - 0\; , \qquad t_c = \frac{1}{8} \; ,
$$
or
\be
C \simeq 1.438 \left ( t_c - t \right )^{1/4} \qquad 
\left ( u = \frac{3}{2}\right ) \; ,
\ee
depending on the ratio $u\equiv U/J$. The ferromagnet--paramagnet phase
transition is of second order for $u>3/2$ and of the first order for
$u<3/2$; the point $u=3/2$ being tricritical, where critical indices
change their values.

The order of the phase transition can be checked by using the Landau
expansion for the free energy
\be
\frac{f}{J} =\left ( w^2 - w +\frac{1}{2}\right ) u + w^2 C^2 -
t\ln\left [ 2{\rm cosh}\left (\frac{w^2C}{t}\right )\right ] \; .
\ee
In our case, we need to substitute here an expansion for the phase
probability (23),
$$
w \simeq \frac{1}{2}\left ( 1 +\frac{C^2}{u} +\frac{C^4}{u^2}\right ) \; .
$$
As a result,
\be
\frac{f}{J} \simeq f_0 + f_2 C^2 + f_4 C^4 \; ,
\ee
where the expansion coefficients are
$$
f_0 =\frac{u}{4} - t\ln 2 \; , \qquad f_2 = -\frac{1-8t}{32t} \; , \qquad
f_4 =\frac{1}{4u}\left ( 3 -\frac{1}{2t} +\frac{u}{768t^3}\right ) \; .
$$
At the critical point $t_c=1/8$, we have $f_2=0$. The criterion of the
second--order transition is $f_4>0$ at $t=t_c$. Since
$$
f_4(t_c) = \frac{2u-3}{12u} \; ,
$$
we should have either $u<0$ or $u>3/2$. The stability conditions (13) and
(15) require that $u\geq 2C^2$. Hence, the second--order transition occurs
for $u>3/2$, while in the interval 
\be
0 < u < \frac{3}{2}
\ee
the transition is of first order. The point $u=3/2$ is tricritical.
And if $u=0$, we return to the standard case of a pure ferromagnet,
without mesophase fluctuations, $w\equiv 1$, with a phase transition at
$t_c=1/2$.

\section{Random--Phase Approximation}

An interesting question, and one of the main for the present paper, is how
the inclusion of magnons changes the results of the mean--field
consideration for a mesophase ferromagnet, especially in the critical
region? To this end, let us introduce the magnon operators [20] for a
$\nu$--replica phase by the equalities
\be
b_{j\nu} = S_{j\nu}^x + i S_{j\nu}^y \; , \qquad
b_{j\nu}^\dagger = S_{j\nu}^x - iS_{j\nu}^y \; .
\ee
These operators have the properties
$$
\left [ b_{i\nu}, b_{j\nu'}\right ] = 0 \; , \qquad b_{i\nu}^2 = 0 \; ,
\qquad
\left [ b_{i\nu}, b_{j\nu'}^\dagger \right ] = \delta_{ij}\delta_{\nu\nu'} 
(1 - 2\hat n_{i\nu}) \; , 
$$
where
$$ \hat n_{i\nu} = b_{i\nu}^\dagger b_{i\nu}
$$ is a magnon density operator in a $\nu$--phase replica. The spin
operators are expressed through the magnon operators as
$$
S_{j\nu}^x = \frac{1}{2}\left ( b_{j\nu}^\dagger + b_{j\nu}\right )\; ,
\qquad
S_{j\nu}^y = \frac{i}{2}\left ( b_{j\nu}^\dagger - b_{j\nu}\right )\; ,
\qquad S_{j\nu}^z =\frac{1}{2} -\hat n_{j\nu} \; .
$$
For a scalar product of two spins, we have
$$
\sS_{i\nu}\cdot \sS_{j\nu} = \frac{1}{4} - \frac{1}{2} \left (
\hat n_{i\nu} + \hat n_{j\nu}\right ) + \hat n_{i\nu}\hat n_{j\nu} +
\frac{1}{2} \left ( b_{i\nu}^\dagger b_{j\nu} +
b_{j\nu}^\dagger b_{i\nu}\right )\; .
$$

Define the magnon propagator
\be
G_{ij\nu}(t) = - i \lgl \hat T b_{i\nu}(t) b_{j\nu}^\dagger(0)\rgl \; ,
\ee
where $\hat T$ is the chronological operator. The random--phase
approximation in the evolution equation for propagator (32) implies that
\be
\lgl b_{i\nu}^\dagger b_{i\nu} b_{j\nu}^\dagger b_{l\nu}\rgl \cong
\lgl b_{i\nu}^\dagger b_{i\nu}\rgl
\lgl b_{j\nu}^\dagger b_{l\nu}\rgl \; .
\ee
Employing the Fourier transforms
$$
G_{ij\nu}(t) =\frac{1}{(2\pi)^4\rho} \int
G_\nu(\sk,\om) \exp\left\{ i (\sk\cdot \sr_{ij} -
\om t )\right \} d\sk d\om \; ,
$$
$$
G_\nu(\sk,w) =\frac{1}{N} \sum_{ij} \int G_{ij\nu}(t) \exp\left \{
-i(\sk\cdot\sr_{ij} - \om t )\right \} dt \; ,
$$
we obtain the equation
\be
\left [ \om - \om_\nu(\sk)\right ] G_\nu(\sk,\om) = 2 S_\nu
\ee
in the random--phase approximation, where the magnon spectrum is
\be
\om_\nu(\sk) = 2 w^2_\nu S_\nu \left [ J - J(\sk) \right ] +
w_\nu B \; .
\ee
The solution of (34) reads
\be
G_\nu(\sk,\om) = 2 S_\nu \left [ \frac{1+n_\nu(\sk)}{\om-\om_\nu(\sk) +i0}
- \frac{n_\nu(\sk)}{\om-\om_\nu(\sk) -i0}\right ] \; ,
\ee
where magnons are treated as bosons and
\be
n_\nu(\sk) = \frac{1}{\exp[\bt\om_\nu(\sk)] -1} = 
\frac{1}{2}\left [ {\rm coth}\frac{\om_\nu(\sk)}{2T} - 1 \right ]\; .
\ee

For an average spin (8) we have
\be
S_\nu = \frac{1}{2} - \frac{1}{N} \sum_j \lgl 
b_{j\nu}^\dagger b_{j\nu} \rgl \; .
\ee
Using the relation
$$
\lgl b_{j\nu}^\dagger b_{j\nu}\rgl = iG_{jj\nu}(-0) =
\frac{i}{(2\pi)^4\rho}\int G_\nu(\sk,\om) e^{+i\om 0} d\sk d\om =
$$
$$
= \frac{2S_\nu}{(2\pi)^3\rho} \int n_\nu(\sk) d\sk \; ,
$$
we obtain the equation
\be
\frac{2S_\nu}{(2\pi)^3\rho} \int {\rm coth}\left [
\frac{\om_\nu(\sk)}{2T}\right ] d\sk =  1 
\ee
for the average spin (38).

For the energy (11), we find
$$
\Sigma_\nu = \frac{J}{4} - \frac{1}{2(2\pi)^4\rho}\int\left [
J - J(\sk) +\frac{\om}{w_\nu^2} -\frac{B}{w_\nu}\right ] e^{-\bt\om}
\times
$$
$$
\times I_\nu(\sk,\om)d\sk d\om \; ,
$$
where the spectral function is
$$
I_\nu(\sk,\om) \equiv -
\frac{2{\rm Im}\; G_\nu(\sk,\om)}{1+e^{-\bt\om}}\; .
$$
Invoking (36), we get
$$
I_\nu(\sk,\om) = \pi S_\nu e^{\bt\om} n_\nu(\sk)\delta\left ( \om -
\om_\nu(\sk)\right ) \; ,
$$ because of which
\be
\sgm_\nu =\frac{J}{4} - \frac{S_\nu(1+2S_\nu)}{(2\pi)^3\rho}\int
\left [ J - J(\sk) \right ] n_\nu(\sk) d\sk \; .
\ee

The spectral representation (36) for the Green function is written
presupposing that magnons are bosons. However, the commutation relations
for magnon operators $b_{i\nu}$ demonstrate that magnons are not strictly
speaking bosons. Magnons can be approximately treated as bosons provided
that the magnon density (the number of magnons per site),
\be
\lgl\hat n_{j\nu}\rgl = \lgl b_{j\nu}^\dagger b_{j\nu}\rgl =
i G_{jj\nu}(-0)\; ,
\ee
is small, being much smaller that $1/2$. This is always correct for a
sufficiently strong external field $B$. But if we are interested in the
case of zero external field $B=0$, then, according to (9) and (10), 
\be
S_1 \equiv C \not\equiv 0 \; , \qquad S_2\equiv 0 \qquad (B=0)\; .
\ee
And from (38) we get
\be
\lgl\hat n_{i2}\rgl =\frac{1}{2} \qquad (S_2 = 0) \; ,
\ee
where the translational invariance of the lattice is assumed. Therefore,
in zero external field, all formulas, starting from (36) and based on this
spectral representation, can be valid only for ferromagnetic phase but not
for paramagnetic one.

In zero external field, the magnon spectrum (35) yields
$$
\om_1(\sk) = 2 w^2 C\left [ J - J(\sk) \right ] \; ,
$$
\be
\om_2(\sk) = 0 \qquad (B = 0)\; .
\ee
The soft mode of magnons in paramagnetic phase tells that, in this phase,
they are in a kind of condensed state.

The phase probabilities $w_1 \equiv w$ and $w_2\equiv 1-w$ are defined by
(12) which, for zero field $B=0$, reads
$$
w = \frac{U - 2\sgm_2}{2(U - \sgm_1 -\sgm_2)} \; .
$$
To make this equation explicit, we need to find the magnetic energies
$\sgm_\nu$ given by (11). For the spin--spin correlation functions, using
the random--phase decoupling (33), with $i\neq j$, we have
$$
\lgl \sS_{i\nu}\cdot \sS_{j\nu}\rgl = \left ( \lgl\hat n_{i\nu}\rgl -
\frac{1}{2}\right )\left ( \lgl\hat n_{j\nu}\rgl -\frac{1}{2}\right ) +
$$
\be
+ \frac{i}{2} \left [ G_{ij\nu}(-0) + G_{ji\nu}(-0) \right ] \; .
\ee
For ferromagnetic phase, we may employ the spectral representation (36)
and get
\be
\lgl\sS_{i1}\cdot\sS_{j1}\rgl = C^2 +\frac{C}{(2\pi)^3\rho}\int
{\rm coth}\left [ \frac{\om_1(\sk)}{2T}\right ] \cos(\sk\cdot\sr_{ij})
d\sk \; ,
\ee
where $i\neq j$. For paramagnetic phase, (45) gives
\be
\lgl\sS_{i2}\cdot\sS_{j2}\rgl = i G_{ij2}(-0) \; .
\ee
To find the Green function for paramagnetic phase, we cannot use (36) but
have to return to the equation of motion leading to (34), which results in
the equation
\be
\om G_2(\sk,\om) = 0 \qquad ( B = 0) \; .
\ee
The momentum dependence of a Green function comes usually through the
spectrum of the corresponding particles. Since the spectrum of paramagnons
does not depend on momentum, it is logical to assume that the related
Green function also does not depend on it. Then the solution to (48) is of
the form
$$
G_2(\sk,\om) = A\delta(\om) \; ,
$$
in which a constant $A$ can be found from conditions (41) and (43). The
latter, together with the equality
$$
\frac{1}{(2\pi)^3\rho} \int\exp(i\sk\cdot\sr_{ij}) d\sk = \delta_{ij} \; ,
$$
where the integration goes over the first Brillouin zone, yield $A=-i\pi$.
Consequently,
\be
G_2(\sk,\om) = -i\pi\delta(\om) \; , \qquad
G_{ij2}(t) = -\frac{i}{2}\delta_{ij} \; .
\ee
Thus, we see that the correlation function (47) is zero for $i\neq j$, and
hence
\be
\sgm_2 =\frac{1}{N}\sum_{i\neq j}J_{ij}\lgl \sS_{i2}\cdot\sS_{j2}\rgl
= 0 \; .
\ee

Now we have all necessary formulas to study thermodynamic characteristics
of the ferromagnet with mesoscopic paramagnetic fluctuations. Let us start
with low temperatures. Then for the ferromagnon momentum distribution
(37), invoking an expansion
$$
{\rm coth} x = 1 + 2\sum_{n=1}^\infty e^{-2nx} \; ,
$$
we may write
$$
n_1(\sk) = \sum_{n=1}^\infty \exp\left \{ - n\; \frac{\om_1(\sk)}{T}
\right \} \; .
$$
For the magnon spectrum, we cam employ an expansion of $J(\sk)$ in powers
of $\sk$ with the averaging over spherical angles and the use of the
equality
$$
\frac{1}{4\pi}\int\cos^n \theta d\Omega = \frac{1}{n+1} \; .
$$
This results in
$$
J(\sk) \simeq J\left ( 1 -\frac{1}{2}a_0^2 k^2 +\frac{1}{40}a_0^4k^4
\right ) \; , \qquad
a_0^2 \equiv \frac{1}{3N}\sum_{i\neq j}\sr_{ij}^2\frac{J_{ij}}{J}\; .
$$
Then, from (44) we find
\be
\om_1(\sk) \simeq JC w^2 a_0^2 k^2 \left ( 1  - \frac{a_0^2k^2}{20}
\right )\; .
\ee

The low--temperature expansion for the average spin becomes
\be
C \simeq \frac{1}{2} - \frac{\zeta(3/2)}{w^3\rho a_0^3}\left (
\frac{t}{2\pi}\right )^{3/2} - \frac{9\pi\zeta(5/2)}{4w^5\rho a_0^3}
\left (\frac{t}{2\pi}\right )^{5/2} \; ,
\ee
where notation (21) is used, and $\zeta(\cdot)$ is a Rieman zeta--function.
For the magnetic energy, we get
\be
\frac{\sgm_1}{J} \simeq \frac{1}{4} - \frac{3\pi\zeta(5/2)}{w^5\rho a_0^3}
\left (\frac{t}{2\pi}\right )^{5/2} -
\frac{45\pi^2\zeta(7/2)}{4w^7\rho a_0^3}
\left (\frac{t}{2\pi}\right )^{7/2} \; .
\ee
It is evident from both (52) and (53) that the presence of mesophase
fluctuations diminishes the average spin and magnetic energy.

The probability of ferromagnetic phase can itself be presented in the form
of a low--temperature expansion resulting from (12),
\be
w \simeq w_0 -\frac{6\pi\zeta(5/2)}{uw_0^3\rho a_0^3}\left (
\frac{t}{2\pi}\right )^{5/2} - \frac{45\pi^2\zeta(7/2)}{2uw_0^5\rho a_0^3}
\left (\frac{t}{2\pi}\right )^{7/2} \; ,
\ee
where $u\equiv U/J$, as in (21), and
\be
w_0 = \frac{2u}{4u-1} 
\ee
is assumed to be nonzero. The same expansion (54) could be obtained from
the equation (7), with the asymptotic form of the free energy
$$
\frac{f}{J} \simeq \left ( w^2 - w +\frac{1}{2}\right ) u -
\frac{w^2}{4} -\frac{2\pi\zeta(5/2)}{w^3\rho a_0^3}\left (
\frac{t}{2\pi}\right )^{5/2} -
$$
\be
- \frac{9\pi^2\zeta(7/2)}{2w^5\rho a_0^3}
\left (\frac{t}{2\pi}\right )^{7/2} \; .
\ee
Combining (56) and (54), we can find expansions for the specific heat
\be
C_V \simeq \frac{15\zeta(5/2)}{4w^3\rho a_0^3}\left (\frac{t}{2\pi}
\right )^{3/2} + \frac{315\pi\zeta(7/2)}{16w^5\rho a_0^3}
\left (\frac{t}{2\pi}\right )^{5/2}
\ee
and the entropy
\be
s \simeq \frac{5\zeta(5/2)}{2w^3\rho a_0^3}
\left (\frac{t}{2\pi}\right )^{3/2} +
\frac{63\pi\zeta(7/2)}{8w^5\rho a_0^3}
\left (\frac{t}{2\pi}\right )^{5/2} \; .
\ee
The stability condition $\prt^2f/\prt w^2 >0$ for the mesophase system to
be stable, at low temperatures, reads
\be
u > \frac{1}{4} +\frac{12\pi\zeta(5/2)}{w^5\rho a_0^3}
\left (\frac{t}{2\pi}\right )^{5/2} +
\frac{135\pi^2\zeta(7/2)}{2w^6\rho a_0^3}
\left (\frac{t}{2\pi}\right )^{7/2} \; .
\ee

It may happen that mesoscopic fluctuations exist only above a temperature
$T_n$ called the nucleation temperature [3]. The latter can be defined by
the condition
$$ 
w = 1 \; , \qquad T = T_n \; ,
$$
leading to an equation $U-2\sgm_1=0$. If the nucleation temperature lies
in the region of low temperatures, then using (53), we obtain
\be
T_n = 2\pi J\left [
\frac{(1-2u)\rho a_0^3}{12\pi\zeta(5/2)}\right ]^{2/5} \; .
\ee
This formula is valid for $u\leq 1/2$.

\section{Phase Transition}

Now pass to the most interesting question, how the interplay 
between mesoscopic and microscopic fluctuations influences the
ferromagnet--paramagnet phase transition? Using the random--phase
approximation, we should keep in mind that the latter is not a good
approximation for the critical region, as well as the mean--field
approximation. Therefore, we do not intend to calculate the critical
indices but rather we pay the main attention to the order of the phase
transition. Our aim is to understand what qualitative changes occur if one
compares two different approximations for a ferromagnet with mesoscopic
fluctuations, the mean--field approximation containing no magnon
fluctuations and the random--phase approximation including magnon
fluctuations.

Consider the critical region, when $C\ra 0$. Then, using an expansion
$$
{\rm coth} x \simeq \frac{1}{x} + \frac{x}{3} - \frac{x^3}{45}
\qquad ( |x| < \pi) \; ,
$$
for the magnon momentum distribution (37), one has
$$
n_1(\sk) \simeq \frac{T}{\om_1(\sk)} - \frac{1}{2} +
\frac{\om_1(\sk)}{12T} - \frac{\om_1^3(\sk)}{720 T^3} \; .
$$

Introduce the magnon factors
\be
m_n \equiv \frac{1}{(2\pi)^3\rho} \int \left [
\frac{J-J(\sk)}{J}\right ]^n d\sk \; ,
\ee
where $n=0,\pm 1,\pm 2,\ldots$. In particular, $m_0=1$. Also, assumimg
that $J_{ii}=0$, one has $\int J(\sk)d\sk=0$. Therefore, $m_1=1$. Denote
\be
m\equiv m_{-1} = \frac{1}{(2\pi)^3\rho}\int
\frac{J}{J-J(\sk)} d\sk > 0 \; .
\ee
The fact that $m>0$ follows from the inequalities
$$
J(\sk) \leq \frac{1}{N}\left | \sum_{i\neq j} J_{ij}\exp\left ( -i\sk
\cdot\sr_{ij}\right ) \right | \leq J \; .
$$
With notation (61), an expansion of (40), as $C\ra 0$, can be written as
$$
\frac{\Sigma_1}{J} \simeq \frac{1}{4} - \frac{t}{2w^2} +
\frac{1}{2}\left ( 1 - \frac{2t}{w^2}\right ) C +
\left ( 1 - \frac{m_2w^2}{6t}\right ) C^2 -
\frac{m_2w^2}{3t} C^3 + \frac{m_4w^6}{90t^3} C^4 \; .
$$

Equation (39) for the average spin yields
\be
3t ( 2mt - w^2 ) + 2 w^4 C^2 - \frac{2m_3w^8}{15t^2} C^4 \simeq 0 \; ,
\ee
as $C\ra 0$. And from the equation (12) for the ferromagnetic--phase
probability, we find
$$
3t ( 2t - w^2 + 4uw^2 - 2 uw ) + 6t ( 2t - w^2 ) C +
2w^2 ( m_2w^2 - 6t ) C^2 +
$$
\be
+ 4m_2w^4 C^3 - \frac{2m_4w^8}{15t^2} C^4 \simeq 0 \; .
\ee
For the critical point, defined by the conditions
$$
t = t_c\; , \qquad w = w_c \; , \qquad C = 0 \; ,
$$ we obtain
\be
t_c = \frac{w_c^2}{2m} \; , \qquad w_c = \frac{2mu}{1+m(4u-1)} \; .
\ee
Introducing the notation
\be
w \equiv w_c + \Delta \; , \qquad \tau \equiv \frac{T - T_c}{T} \; ,
\ee
we may rewrite expansions (63) and (64) as follows:
$$
\Delta \simeq \frac{w_c}{2m(4u-1)} \left\{ - \tau + \left [
\frac{w_c^2(-\tau)}{m(4u-1)} + 2( m - 1 )\right ] C + \right.
$$
$$
\left. + \left [ \frac{2w_c^2(m-1)}{m(4u-1)} +
\frac{4m(3 - mm_2w_c^2)}{3w_c^2} + \frac{w_c^5(-\tau)}{m^2(4u-1)^2} -
\frac{4m_2w_c^2(-\tau)}{3(4u-1)}\right ] C^2 \right \} \; ,
$$
and
$$
\frac{3mu(-\tau)}{w_c^3} + \frac{3}{2m}\left [ m - 1 +
\frac{w_c^2(-\tau)}{2m(4u-1)}\right ] C +
$$
$$
+ \left [ \frac{3}{w_c^2} - m( 4u - 1 ) - mm_2 +
\frac{3w_c^2(m-1)}{2m^2(4u-1)} + \frac{3(-\tau)}{m(4u-1)^2} -
\frac{m_2w_c^2(-\tau)}{m(4u-1)}\right ] C^2 \simeq 0 \; .
$$
From here, two types of critical behaviour can arise depending on whether
the factor $m$ is equal to one or not. In the first case, we get
\be
C \simeq 2 \left ( \frac{6u}{4u-13+m_2}\right )^{1/2} (-\tau)^{1/2}
\qquad (m=1)
\ee
for the average spin, and for $\Delta\equiv w-w_c$, we find
\be
\Delta \simeq \frac{32u(12-m_2)}{4u-13+m_2}(-\tau) \qquad (m=1)\; .
\ee
In the case $m\neq 1$, we obtain
\be
C \simeq \frac{2um^2}{w_c^3(m-1)} \tau \qquad (m\neq 1)
\ee
for the order parameter, and also
\be
\Delta \simeq \frac{w_c^3 - 4m^2u}{2mw_c^2(4u-1)}(-\tau) \qquad
(m\neq 1) \; .
\ee

The case $m=1$ leads to formulas (67) and (68) with a mean--field type
behaviour, provided that $4u-13+m_2>0$. In order to decide what kind of
behaviour would correspond to reality, let us take the values of the
parameters $m$ and $m_2$ for several crystalline lattices [20]. For the
simple cubic lattice, one has $m=1.516$ and $m_2=1.167$. For the
body--centered cubic lattice, one gets $m=1.393$ and $m_2=1.125$. And for
the face--centered cubic lattice, one has $m=1.345$ and $m_2=1.083$.
Stability conditions (13) and (15) require that
$$
u > \frac{m-1}{2m} \qquad ( t = t_c )
$$
at the critical point. For a simple cubic lattice this gives $u>0.170$;
for b.c.c. lattice, $u>0.141$; and for f.c.c. lattice, $u>0.128$. Since
$m\neq 1$, we have to consider formulas (69) and (70). But for $m>1$, the
order parameter (69) approaches zero at the critical point (65) from the
right side of $t_c$. This means that such a critical point, actually, does
not exist [3--5], but there is a temperature $t_0>t_c$ where the first
order phase transition occurs.

\section{Conclusion}

We have considered here a model of a ferromagnet with mesoscopic
paramagnetic fluctuations. The ferromagnet with such mesophase
fluctuations can be thermodynamically more stable than a pure ferromagnet.
A necessary condition for the appearance of mesophase fluctuations is a
sufficiently strong crystal field.

The occurrence of mesoscopic fluctuations in a ferromagnet leads to a
lower, as compared to the case of a pure ferromagnetic phase, magnetization. 
The spin--wave spectrum softens, the spin--wave attenuation increases,
and the magnon free path shortens. The laws characterizing these effects
can be expressed by the relations
$$
M \sim w\; , \qquad \ep(\sk) \sim w^2 \; , \qquad
\gm(\sk) \sim w^{-3} \; , \qquad \lmb(\sk) \sim w^5 \; ,
$$ in which $w\leq 1$ is the probability of ferromagnetic phase.

In the mean--field approximation, the order of the ferromagnet--paramagnet
phase transition in the mesophase ferromagnet depends on the ratio
$u\equiv U/J$ of the crystal--field parameter $U$ to the exchange
interaction $J$. For $u\leq 0$, there are no mesophase fluctuations, and
the standard second--order transition happens. For $u>0$, mesophase
fluctuations exist in some temperature region above the nucleation
temperature. In the interval $0<u<3/2$, the phase transition becomes of
first order. The value $u=3/2$ corresponds to a tricritical point. And for
$u>3/2$, the phase transition in the mesophase ferromagnet is again of 
second order.

In the random--phase approximation, mesophase fluctuations can exist in
the critical region for $u>(m-1)/2m$, where the magnon factor $m$ depends
on the kind of crystalline lattice, ranging from $m=1.52$ for a simple
cubic lattice to $m=1.35$ for a f.c.c. lattice. At low temperatures,
mesophase fluctuations can arise if $u>1/4$. The phase transition, for the
realistic values of the magnon factor $m> 1$, becomes of first order for
any $u$ satisfying stability conditions.

\vspace{5mm}

{\bf Acknowledgement}

\vspace{2mm}

I am very grateful for useful remarks to G.A. Gehring, D. ter Haar, H.
Hasegawa, R.B. Stinchcombe, and I.R. Yukhnovsky, whom the early variants
of this work have been discussed with. The Senior Fellowship from the
University of Western Ontario, London, Canada is appreciated.

\end{document}